# A new type of carbon resistance thermometer with excellent thermal contact at millikelvin temperatures

Nodar Samkharadze, Ashwani Kumar, and Gábor A. Csáthy

*Department of Physics, Purdue University, 525 Northwestern Avenue, West Lafayette, IN 47907-2036, nsamkhar@purdue.edu*

Using a new brand of commercially available carbon resistor we built a cryogenic thermometer with an extremely good thermal contact to its thermal environment. Because of its superior thermal contact the thermometer is insensitive to low levels of spurious radio frequency heating. We calibrated our thermometer down to 5mK using a quartz tuning fork He-3 viscometer and measured its thermal resistance and thermal response time.

Resistance thermometry is one of the simplest ways of measuring temperature. For cryogenic temperatures below 1K there are numerous examples of resistance thermometers[1] such as thick film ruthenium oxide (RuO), Cernox, doped germanium, and various carbon composition resistors such as Allen-Bradley, Speer, and Matsushita. While these thermometers work reliably over wide temperature ranges, it is well known that there are difficulties in using them below a few tens of millikelvin[1]. At these temperatures the resistance is often independent of temperature or irreproducible over time. Such a behavior is thought to be due to poor thermal contact of the thermometer to its surroundings. Indeed, minute radio frequency pickup or ground loop current causes unintentional heating of the sensor above the phonon temperature of its environment. Such a heating results in a temperature reading which is higher than that of the bath. Consequently resistance thermometry is not commonly used at the lowest temperatures of a few mK achieved by today's dilution refrigerators. Among the resistance thermometers only carbon composition resistors of the Speer[2,3,4,5] and Matsushita[5] brands are known to work below 20mK. These two kinds of resistors have been used as secondary thermometers from the early 60s, but unfortunately they are long out of production and today can only be obtained from stocks of a few generous people. We know of a single commercially available Speer thermometer mounted in a metal housing[6].

In this paper we report on a resistance thermometer which responds to temperatures as low as 5mK and which is built using a previously unknown brand of carbon composition resistor, Little Demon of the Ohmite Manufacturing Co[7]. Among the known resistance thermometers, our thermometer has the most favorable thermal contact at mK temperatures and therefore is the least susceptible to spurious ground loop and radio frequency heating. The latter type of heating often renders thermometers unusable at mK temperatures, it is notoriously difficult to control, and it is setup dependent. In order for a resistance thermometer to retain its calibration in the mK temperature range when moved to different refrigerators, it must have a good thermal contact to its environment. Our thermometer is the best in this regard since its thermal contact at 6mK is 5 times better than that of Speer thermometers[3,4] and 200 times better than the extrapolated values for RuO thermometers[8,9]. In addition, our thermometer has the convenience of reasonable thermal response times and of a monotonic response from 5mK all the way to room temperature.

Details of the thermometer which we will refer to as Th-1 are shown in the inset of Fig.1. In order to achieve a good thermal contact of the carbon sensor to the copper housing we adapted a well known techique[3,4,10]. The sensor is made of a 47Ω 1/2W resistor with 5% tolerance[11]. The phenolic packaging of the resistor was removed and the resulting carbon cylinder was ground down to a 0.3 mm thick rectangle at which point its resistance at room temperature became 5.7 times larger than its original value. As seen in the inset of Fig.1, insulated copper wires of AWG 32 are soldered on with Indium to the remains of the original copper leads of the resistor which are kept embedded in the carbon. Such a design is known to yield superior electrical contact to the sensing element[10]. We chose Indium solder because using commonly available lead solder caused a 12% decrease of the resistance as measured at room temperature. We note that such a change of resistance is consistent with results of heating of Speer resistors in ambient atmosphere[12]. When using Indium, we do not find any measurable change after soldering. As an alternative to Indium we suggest the use of conductive silver epoxy. The thinned resistor was inserted in the 0.5mm slot of the copper housing of 99.99% purity[3,5]. As seen in the inset of Fig.1, the slot has two drilled holes which are used to accommodate the wires and the solder joints of the sensor. To maintain electrical isolation we used two sheets of 7.6 μm thick Kapton foil[13] and the pieces were secured together by Stycast 1266 epoxy. The two copper leads were tightly wrapped around and attached to the housing with the same type of epoxy. In order to avoid well known calibration shifts with repeated cool-downs we made sure the sensing element is hermetically sealed with epoxy and the thermometer has been thermally cycled 10 times to 77K and 6 times to 4.2K.

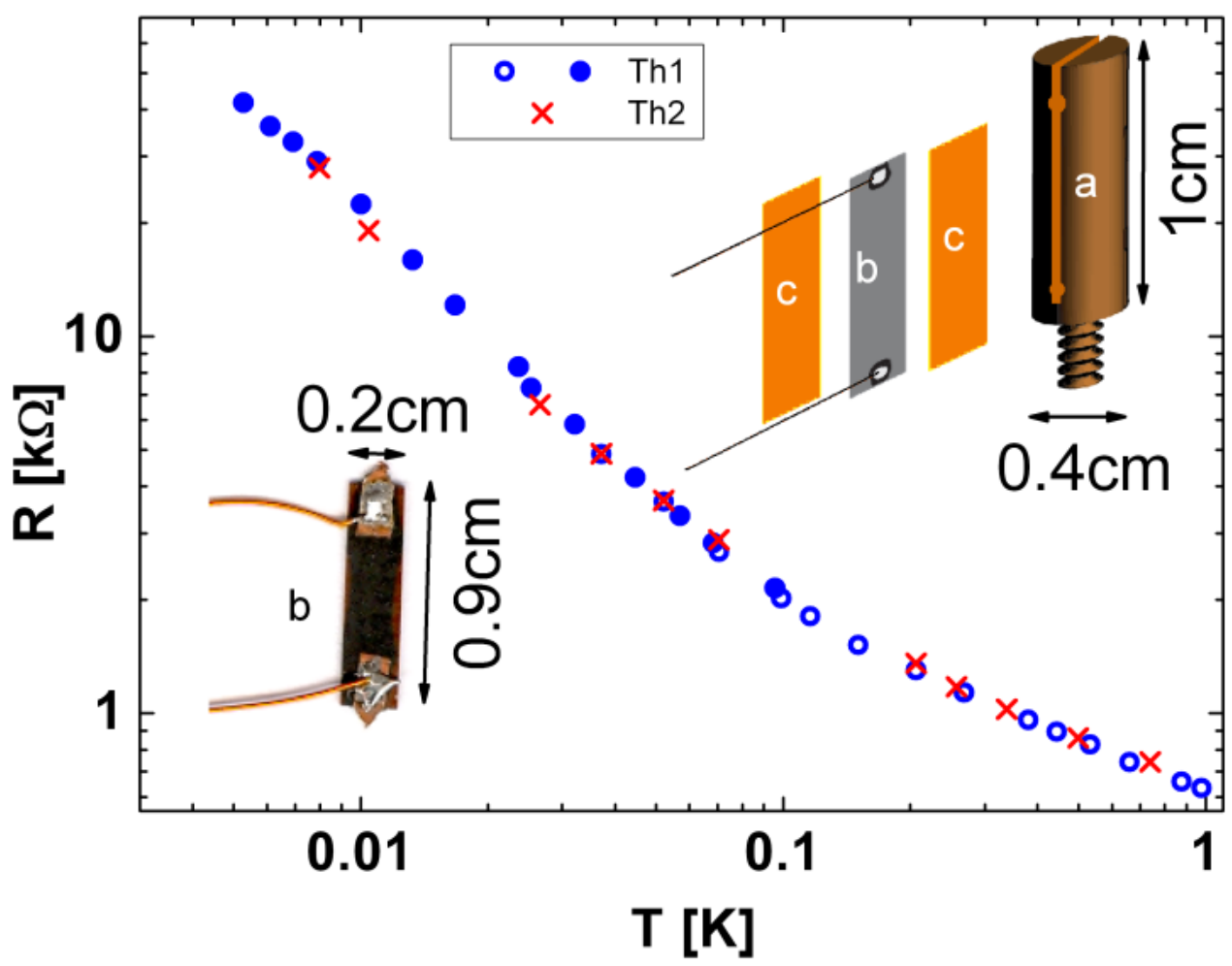

FIG.1. *Resistance of the two carbon thermometers R versus temperature T. Open and close circles show the cross-calibration of Th-1 against a commercial RuO thermometer and a quartz tuning fork He-3 viscometer, respectively. The two calibration curves for Th-1 overlap between 44 and 100 mK. The crosses are the calibration by similar means of a second carbon thermometer Th-2 which we immerse directly into He-3 liquid. Insets shows details of the thermometer a:Cu housing, b:carbon sensing element with leads attached, c: Kapton sheet.*

The thermometer was screwed into the mixing chamber of our dilution refrigerator and was cross-calibrated against three other thermometers: a commercial silicon diode at high temperature, a calibrated RuO thermometer down to 44mK[14], and a homebuilt quartz tuning fork viscometer immersed into He-3 below 100mK. Our quartz viscometer is similar to the one recently studied[15], its details will be described below. Besides Th-1 we prepared a second thermometer Th-2 from the same batch of 47 Ω 1/2W resistors. Th-2 consists of the bare carbon sensor as seen in the lower left corner of Fig.1 which is *not* glued into a copper housing but which is instead immersed directly into the liquid He-3 in close proximity to our quartz viscometer.

Filters on the electrical leads are critical in any resistance measuring setup for low temperatures. We used two different ways of radio frequency filtering. The first one is an rf filter at room temperature and it consists of a commercial filter housed in a D-sub connector[16]. The connector is secured inside an aluminum box. This box has a second connector which is plugged in directly into the manufacturer provided leak-tight Fischer connector mounted on the top of the cryostat. The second type of filter consists of copper magnet wires AWG 32 embedded into conductive silver epoxy[17] at low temperature. The silver epoxy also serves as glue which attaches and thermally anchors the wires to the surface of the foot long copper tail fastened to the mixing chamber. We have not characterized our transport setup without the two sets of rf filters. With the filters in place, the electromagnetic environment of our resistance thermometers can be characterized by an rf heating power of 100aW or less, as is discussed later in the paper. We found that the installation of a conventional RC filter of 50kHz cutoff frequency on the Still did not have a measurable effect. Four-lead resistance measurements of the thermometers in the filtered setup described using a lock-in amplifier and a LakeShore 370 resistance bridge with a model 3716 scanner front end gave consistent readings.

We turn our attention to the He-3 viscometer used for calibration below 100mK. He-3 viscometers are useful in measuring temperature since the viscosity $\eta$ of He-3 liquid is a strong function of the temperature *T*. According to the Fermi liquid theory $\eta$ obeys the simple equation $\eta T^2$=constant in the strongly degenerate limit, below about 30mK[18]. At higher temperatures corrections must be included[19]. Our refrigerator is already equipped with a He-3 cell with a sintered silver heat exchanger used for cooling two-dimensional electron gases in semiconductor quantum wells[20]. The viscometer is thus a natural choice for measuring *T*. In addition, viscometers have a number of advantages: they are immune to radio frequency heating, are simple as compared to melting curve or paramagnetic susceptibility thermometers used in this temperature range, dissipate very little heat, have a relatively quick response time, and have practically no dependence on the magnetic field. To find *T* we measure the quality factor of an oscillating quartz tuning fork immersed into the liquid He-3. The quartz crystal resonant frequency is 32.768kHz. Its can and original leads have been removed as they are ferromagnetic. Recent work[15] on a similar viscometer based on a quartz tuning fork of the same physical size as ours confirmed the expected behavior of the quality factor dependence on *T* at mK temperatures. The inset of Fig.2 shows a typical Lorentzian line shape of the in-phase $I_x$ and out-of-phase $I_y$ components of the resonant currents we measure through the quartz at 10mK as function of the driving frequency at 5mV excitation voltage.

Fig. 1 displays the variation of the resistance measured in the limit of no self-heating below 1K for both carbon thermometers. The resistance at the lowest temperature does not exceed 50kΩ which can be comfortably measured with our electronics. The thermometer Th-1 in the 0.01 to 1K range has a smooth response which can be fitted with the variable range hopping equation $R(T) = R_0 \exp(T_0/T)^{\alpha}$, where $R_0$=211Ω, $T_0$=1.39K, and the exponent α=0.31±0.01. This fitting equation has been used for other carbon composition thermometers[4,21,22]. Above 1K, in contrast to RuO thermometers, both our thermometer responds to changes in temperature all the way to room temperature, albeit with a progressively decreasing sensitivity. At around 10mK there is an inflexion point in the calibration of Th-1 which is puzzling. One possibility for the origin of this inflexion point is an error of our quartz viscometer temperature reading. We argue that this is not likely since recent results on cross-calibrating a quartz tuning fork of the same physical dimensions as ours against a melting curve thermometer do not exhibit any curvature in the calibration around 10mK[15]. In a subsequent paragraph we find that the spurious radio frequency heating in our setup is insufficient to cause a self-heating of our thermometer and therefore it cannot explain the inflection around 10mK. We thus surmise that the inflexion around 10mK is an intrinsic property of our thermometer rather than a result of measurement artifacts and it represents a change from a hopping type of conduction to a different mechanism of conduction. Such a crossover between two different conduction regimes of the *R* versus *T* calibration is not unique to the carbon resistor we use. A Matsushita resistor shows a crossover from a hopping law with exponent α=½ to another one with α=¼ in the vicinity of 300mK[21] and in a recent paper it is suggested that Speer thermometers exhibit a similar crossover close to 90mK[22]. We note that 9 repeated thermal cycling of Th-1 from room temperature to 5mK caused changes of the calibration of 2% as measured against a RuO thermometer and the quartz viscometer. Fig.1 also shows that the calibration of Th-1 and Th-2 coincides to a good extent. While the resistors used in these thermometers come from the same batch, the coincidence is unexpected as the room temperature resistance is 268Ω and 281Ω for Th-1 and Th2, respectively. We conclude that thermometers of similar room temperature resistance have a spread in their calibration curves and for reliable measurements therefore each thermometer has to be individually calibrated.

To assess the thermal contact of the thermometer Th-1 to its environment, we perform a self-heating experiment. We control the temperature of the mixing chamber $T_{mix}$ and apply successively larger excitation currents. A higher excitation will heat the thermometer above the temperature of the mixing chamber. This heating result in a lower resistance and using the calibration done at low excitation shown in Fig.1 we convert the resistance of the heated thermometer into temperature. Several such self-heating curves at different mixing chamber temperatures are shown in Fig.2. We notice that in order to heat Th-1 to 50mK when the surrounding is at the base temperature we need a 2.2nW excitation power. For the same self-heating to 50mK RuO thermometers require 10pW[8] or 30pW[9].The 100 fold larger power required for a same self-heating suggests a superior thermal contact of Th-1 with its environment as compared to that of RuO thermometers.

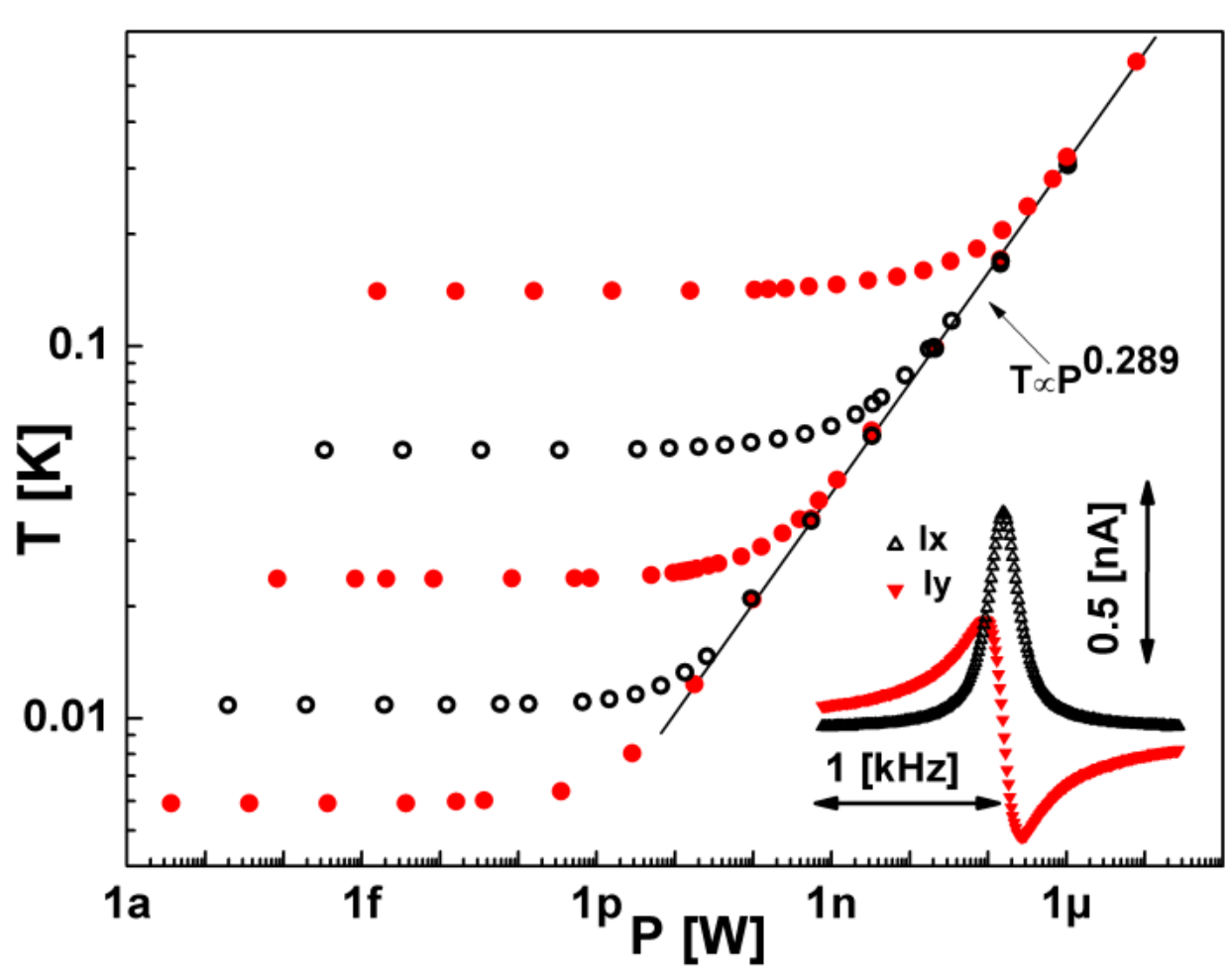

FIG.2. *Self-heating curves T vs the Joule excitation power P of thermometer Th-1 at several different mixing chamber temperatures. Curves have a common asymptotic limit at high heating powers shown by a straight line. Inset shows the in-phase and out-of-phase components of current through the quartz resonator as function of the frequency at 10mK.*

The strength of the thermal contact of a thermometer to its environment is measured by the thermal resistance $R_{th}$ defined as an infinitesimal temperature change due to a small self-heating excitation power. As shown in the inset of Fig.3 for $T$=29.2mK, $R_{th}$ is the slope of the self-heating curves when plotted on a linear scale. The points at $P$<0.15pW are not shown in the inset as they have an error in $T$ due to amplifier noise larger than the typical separation of ~0.03mK between the points and hence they are not useful in the determination of the slope. The thermal resistances for Th-1obtained this way at various temperatures are plotted in Fig. 3.

Before comparing the numerical values to those of other types of thermometers we note that a lower $R_{th}$ is desirable as it translates into a better thermal contact to the bath. We find that above 80mK the thermal contact of our thermometer is comparable to that of RuO[23], Speer[24], and Ge[23] thermometers and it is superior when compared to Cernox[23] and another RuO[8]. Below 30mK, however, our thermometer has the best thermal contact. At 6mK $R_{th}$ is 1.5x10$^{9}$K/W for our thermometer as compared to 3.3x10$^{10}$K/W[3] and 4.0x10$^{10}$K/W[4] for Speer thermometers. Hence the thermal contact at 6mK of our thermometer Th-1 is about a factor of 5 better than that of Speer thermometers[3,4] and it is at least a factor of 200 better than the extrapolated value for RuO[8,23]. We conclude that at the lowest temperatures our thermometer has the best thermal contact to its environment among all resistance thermometers. This property makes our thermometer the least susceptible to heating from its electromagnetic environment and allows measurements with negligible self-heating at higher excitation levels. The power levels which cause a 1% increase of the temperature of the thermometer are 1.5pW at 29.2mK, 120fW at 11.5mK, and 48fW at 6.5mK.

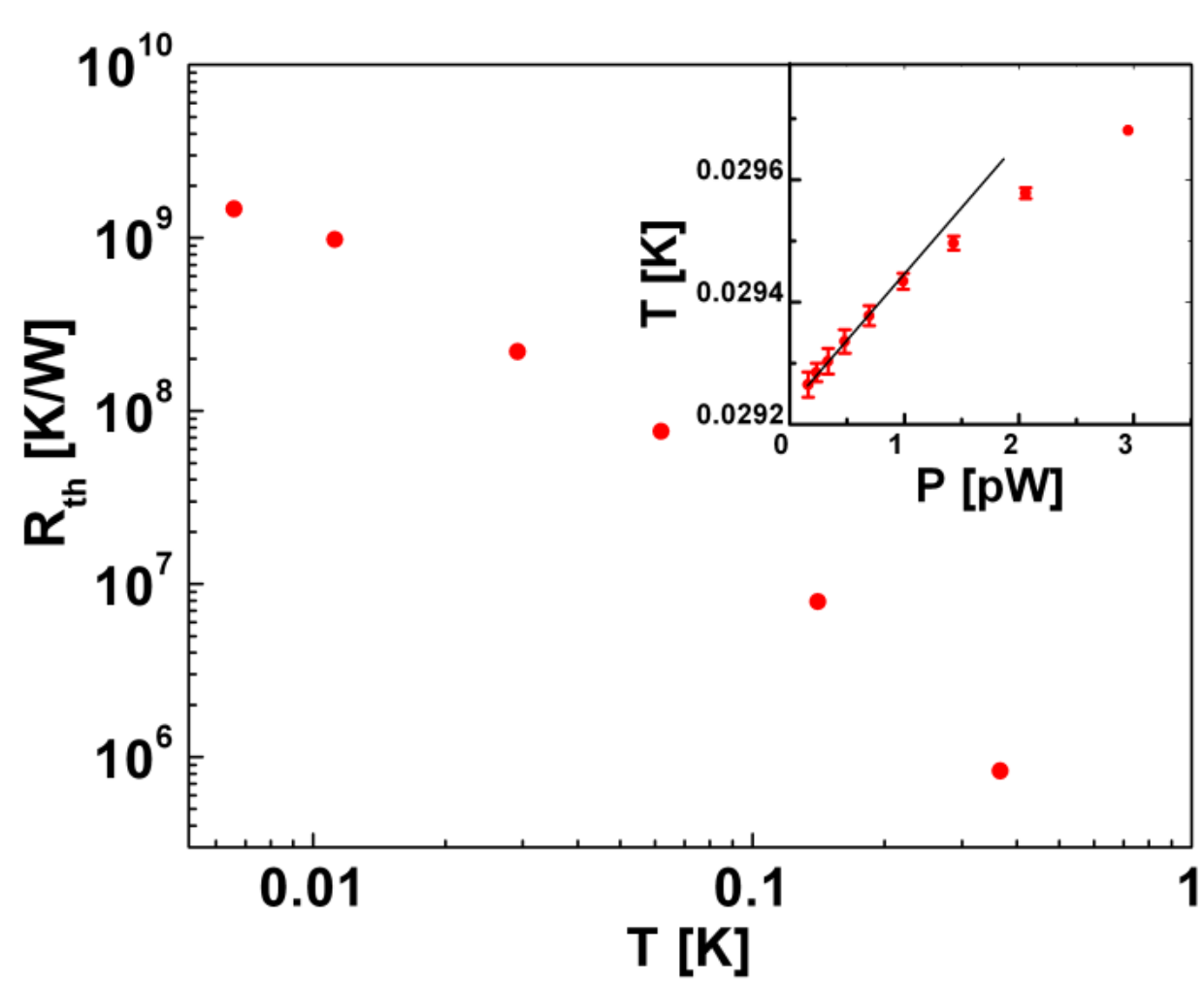

FIG.3. *Thermal resistance $R_{th}$ as function of the temperature T for Th-1. The inset illustrates the extraction of the thermal resistance from the slope of the self-heating curve at $T_{mix}$=0.0292K.*

The thermal resistance of Speer thermometers was inferred to be dominated by that of the carbon sensor itself[24]. We reached the same conclusion by examining our second carbon sensor Th-2 which was soldered onto sintered silver heat exchangers[20] and was directly immersed into the liquid He-3. Self-heating curves at the base temperature of the fridge are found to be identical within measurement error for the two carbon thermometers we made (not shown). The conclusion above is further strengthened by a remarkable observation that the thermal resistance of graphite calculated for the geometry of our thermometer using the extrapolated values of the thermal conductivity measured above 0.3K[25], is within a factor of 2 from that of Th-1 over the whole $T$ range.

A short thermal response time is another desirable property of a thermometer. Since a quick change in the temperature of the mixing chamber is not possible, we extract the temperature dynamics by self-heating the thermometer. We used a dc self-heating current which was disconnected at the time $t$=0 while a small ac excitation was left on to monitor the resistance. The inset to Fig. 4 shows the temporal response of the temperature difference after heating the thermometer Th-1 2.5mK above the mixing chamber temperature $T_{mix}$. Since the thermal resistance and heat capacity of the thermometer are not independent of the temperature, the curves of the inset do not follow a simple exponential decay. We define therefore the thermal response time $\tau$ as the time interval necessary for the temperature difference to drop to 10% of its initial value at $t$=0, i.e. the time necessary for the temperature $T$ to be within 0.25mK from $T_{mix}$. We find that at 20mK $\tau$ is too short to be measured with our electronics and that $\tau$ increases rapidly with decreasing $T$. We have not observed any relaxation of long timescales reported for a Speer thermometer[26]. The measured relaxation times are consistent with the results from switching the Lakeshore model 3716 scanner between our carbon and a RuO thermometer. At such a switching event we think there is a small transient in the excitation which heats up the connected thermometer momentarily after which the resistance reading settles. We observe that the settling time after switching for our thermometer is more than a factor ten shorter than that for a RuO thermometer.

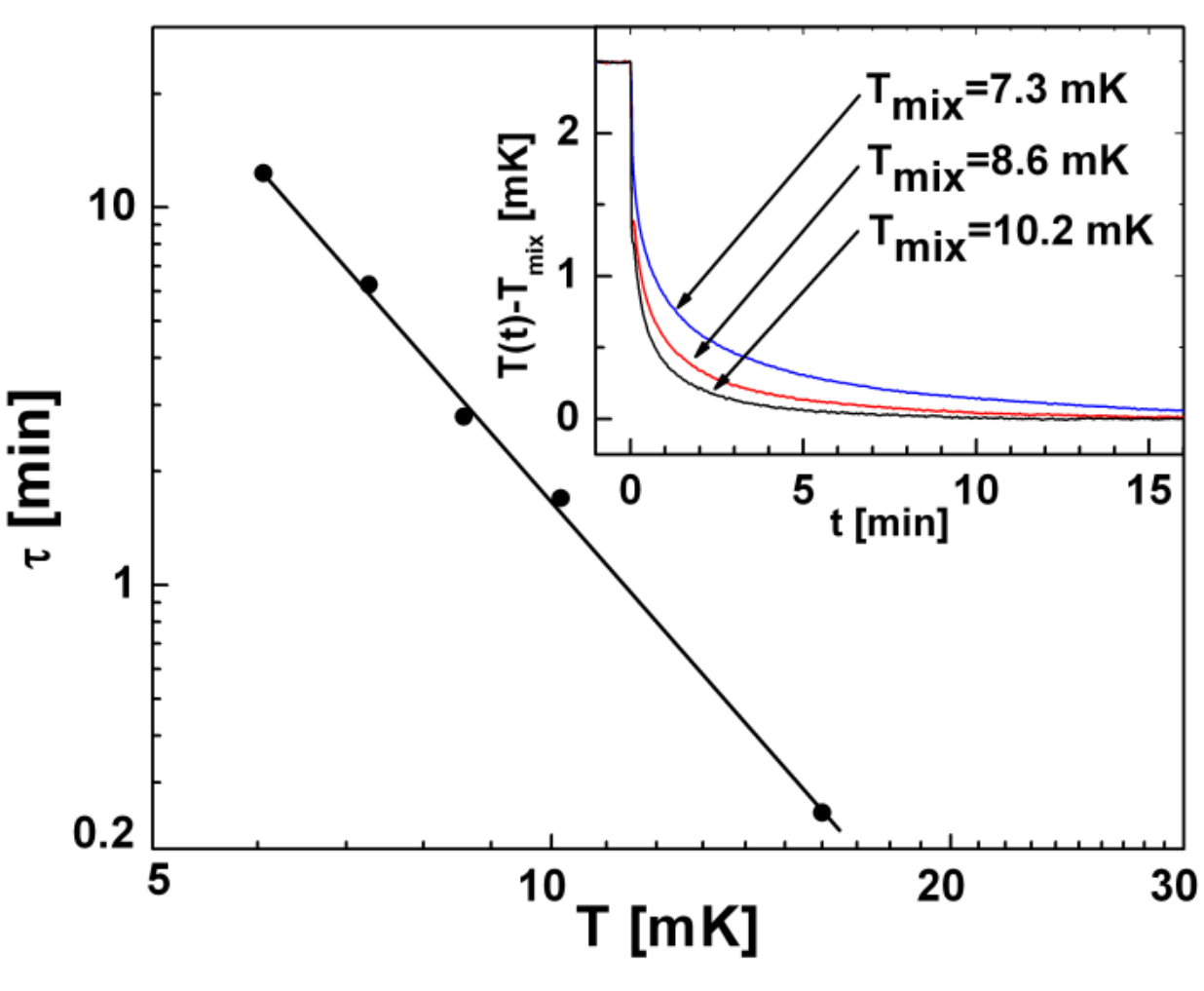

FIG.4. *Thermal response time $\tau$ vs T for Th-1. The inset shows the temperature relaxation after heating the thermometer 2.5mK above the mixing chamber temperature $T_{mix}$.*

We mentioned earlier that spurious radio frequency heating is present in all instruments due to the presence of the measuring leads. Upper limits of the radio frequency heating powers can be estimated from self-heating measurements. However, our carbon thermometers are not useful in estimating this spurious power because of their strong thermal contact to the bath. Therefore we choose a resistive thermometer with a large thermal resistance which responds dramatically to low levels of heating, a RuO thermometer. We estimate that the residual power level the measuring leads bring in is of the order of 100aW or less for our refrigerator. We note that such a power level does not cause any measurable self-heating in our carbon thermometer.

In conclusion, we built a carbon resistor based thermometer which can be routinely used as a secondary thermometer down to 5mK provided good grounding and shielding practices are followed. The carbon resistor is commercially available and can be used to replace the obsolete Speer resistors no longer in production. The thermal resistance at 6mK of our thermometer is a factor 5 better than that of Speer thermometers and a factor of 200 better than that of RuO thermometers. This means that among resistance thermometers at mK temperatures our thermometer has a superior thermal contact to the bath and therefore it has good immunity to spurious heating due to ground loop and radio frequency heating. The good thermal contact also results in reasonable thermal response times even at the lowest temperatures.

## Acknowledgements

We acknowledge useful discussions with G.Frossati and J.S.Xia. N.S. and G.A.C. were supported on NSF grant DMR-0907172.